\documentclass[preprint,showpacs,preprintnumbers,amsmath,amssymb]{revtex4}
\usepackage{amssymb}
\usepackage{latexsym}
\usepackage{epsfig}                                   
\usepackage{amsmath}

\begin{document}

\title{Thermodynamical study of FRW universe in Quasi-Topological Theory}
\author{H. Moradpour\footnote{h.moradpour@riaam.ac.ir} and R. Dehghani}
\address{Research Institute for Astronomy and Astrophysics of Maragha (RIAAM), P.O. Box 55134-441, Maragha, Iran}
\begin{abstract}
By applying the unified first law of thermodynamics on the apparent
horizon of FRW universe, we get the entropy relation for the
apparent horizon in quasi-topological gravity theory. Throughout the
paper, the results of considering the Hayward-Kodama and Cai-Kim
temperatures are also addressed. Our study shows that whenever,
there is no energy exchange between the various parts of cosmos, we
can get an expression for the apparent horizon entropy in
quasi-topological gravity, which is in agreement with other attempts
followed different approaches. The effects of a mutual interaction
between the various parts of cosmos on the apparent horizon entropy
as well as the validity of second law of thermodynamics in
quasi-topological gravity are also perused.
\end{abstract}
\pacs{04.20.-q, 04.50.-h, 98.80.-k, 95.35.+d, 95.36.+x}
\maketitle

\section{Introduction\label{Intr}}
Since observational data indicates an accelerating universe
\cite{Rie,Rie1,Rie2,Rie3}, one should either consider a non-trivial
fluid in the Einstein relativity \cite{Rev3,Rev1,Rev2} or modifying
the Einstein theory \cite{Rev1,mod,meeq,lobo}. Moreover, there are
some observational evidences which permit a mutual interaction
between the dark sides of cosmos
\cite{z1,z2,z3,pavonz,z4,z5,z6,z7,z8,wangpavon,z9,co1,z10,co2,co3},
including the dark energy and dark matter \cite{roos}. The
thermodynamic consequences of such interactions in the Einstein
framework are addressed in Ref.~\cite{jss}. Recently, it is argued
that since the origin of dark energy is unknown, some dark energy
candidates may affect the Bekenstein entropy of apparent horizon in
a flat FRW universe \cite{cana,cana1,mmg,mms}. In addition, by
following \cite{em}, one can see that all of the dark energy
candidates and their interaction with other parts of cosmos may
affect the apparent horizon of a flat FRW universe in the Einstein
framework. There are additional terms to the Einstein tensor in
modified theories of gravity. Since these additional terms can be
interpreted as a geometrical fluid, a mutual interaction between the
dark sides of cosmos may be an interaction between these geometrical
and material terms \cite{lobo}. Therefore, one may expect that such
modifications to Einstein theory and their interaction with other
parts of cosmos may also affect the horizon entropy. Indeed, there
are some new attempts in which authors give a positive answer to
this expectation and show that such geometrical fluids and their
interaction with other parts of cosmos may also affect the horizon
entropy of FRW universe \cite{mitra,mitra1,mitra2,mitra3}.

String theory together with AdS/CFT correspondence conjecture give
us the motivation to study spacetimes with dimensions more than
$4$ \cite{Pol,Mad,Mad1}. Moreover, brane scenarios also encourage
us to study $(4+1)$-dimensional spacetimes \cite{RS1,RS}. The
backbone of Einstein-Hilbert action is producing the second order
field equations of motion. In looking for a suitable lagrangian
for higher dimensions, which also keeps the field equations of
motion for metric of second-order, one reaches to Lovelock
lagrangian instead of the usual $(n+1)$ version of Einstein-Hilbert
action \cite{lov}. It is shown that the corresponding
gravitational field equations of static spherically symmetric
spacetimes in the Einstein and Lovelock theories are parallel to
the availability of the first law of thermodynamics on the
horizons of considered metric \cite{PSP}. For a FRW universe,
where the field equations governs the universe expansion history,
one can also reach at the corresponding Friedmann equations in the
Einstein and Lovelock theories by applying the first law of
thermodynamics on the apparent horizon of FRW universe
\cite{Caikimt}. Indeed, such generalization of Einstein-Hilbert
action leads to additional terms besides the Einstein tensor in
spacetimes with dimensions more than $4$. Authors in \cite{mitra}
interprets such additional terms as the geometrical fluids and
show that these terms and their interactions with other parts of cosmos affect the horizon entropy.

By considering cubic and higher curvature interactions, some authors
try to find new lagrangian which keeps the field equation of motion
for the metric of second-order \cite{oliv,Myer1,MHD2}. Their theory
is now called quasi-topological gravity which affects the
gravitational field equations in spacetime with dimensions more than
$4$. Thermodynamics of static black holes are studied in this theory
which leads to an expression for the horizon entropy
\cite{oliv,Myer1,MHD2}. Thermodynamics of static black holes are
studied in this theory which leads to an expression for the horizon
entropy \cite{Myer1,MHD2,MHD3,Brenn,DSH}. Moreover, it is shown that
if one assumes that the black hole entropy relation, derived in
\cite{Myer1,MHD2,MHD3,Brenn,DSH}, is available for the apparent
horizon of FRW universe, then by applying the unified first law of
thermodynamics on the apparent horizon of FRW universe, one can get
the corresponding Friedmann equations in quasi-topological gravity
\cite{paper1}. Thereinafter, authors point out the second law of
thermodynamics and its generalized form \cite{paper1}. In fact, this
scheme proposes that, in quasi-topological gravity, one may
generalize the black hole entropy to the cosmological setup by
inserting the apparent horizon radii instead of the event horizon
radii. Is it the only entropy relation for the apparent horizon?
Does a mutual interaction between the geometrical fluid, arising
from the terms besides the Einstein tensor in the field equations,
and other parts of cosmos affect the horizon entropy?

Our aim in this paper is to study the thermodynamics of apparent
horizon of FRW universe in the quasi-topological gravity theory to
provide proper answers for the above mentioned questions. In this
investigation, we point to the results of considering the
Hayward-Kodama and Cai-Kim temperatures for the apparent horizon. In
order to achieve this goal, by looking at the higher order curvature
terms, arising in quasi-topological gravity, as a geometrical fluid
and applying the unified first law of thermodynamics on the apparent
horizon, we get an expression for the apparent horizon entropy in
this theory, while the geometrical fluid does not interact with
other parts of cosmos. Moreover, we focus on a FRW universe in which
the geometrical fluid interacts with other parts of cosmos. This
study signals us to the effects of such mutual interaction on the
apparent horizon entropy of a FRW universe with arbitrary curvature
for quasi-topological gravity theory. As the non-interacting case,
the second law of thermodynamics is also studied in an interacting
universe.

The paper is organized as follows. In the next section, we give an
introductory note about the quasi-topological gravity, the
corresponding Friedmann equations and some properties of FRW
universe, including its apparent horizon radii, surface gravity and
thus the corresponding Hayward-Kodama temperature.
Section($\textmd{III}$) devotes to a breif discussion about the
Unified First Low of thermodynamics (UFL). In section
($\textmd{IV}$), bearing the Hayward-Kodama temperature together
with the unified first law of thermodynamics in mind, we study the
effect of a geometrical(curvature) fluid on the apparent horizon
entropy in a FRW universe with arbitrary curvature parameter for
Quasi-topological theory. The results of considering the Cai-Kim
temperature are also addressed. Thereinafter, we generalize our
study to a universe in which the geometrical fluid interacts with
other parts of cosmos. The results of attributing the Cai-Kim
temperature to the apparent horizon in an interacting universe are
also pointed out in section ($\textmd{IV}$). Section ($\textmd{V}$)
includes some notes about the availability of second law of
thermodynamics in an interacting cosmos described by
quasi-topological gravity theory. The final section is devoted to
summary and concluding remarks.
\section{A Breif Overview of Quasi-Topological Gravity\label{Quasi1}}
A natural generalization of the Einstein-Hilbert action to higher
dimensional spacetime, and higher order gravity with second order
equation of motion, is the Lovelock action
\begin{equation}\label{Act1}
I_{\mathrm{G}}=\frac{1}{16\pi G_{n+1}}\int d^{n+1}x\sqrt{-g}\left(\sum^{m}_{i=1} c_i\mathcal{L}%
_{i}+\mathcal{L}_{\rm M}\right).
\end{equation}
where $c_i$'s are Lovelock coefficients, $\mathcal{L}_i$'s are
dimensionally extended Euler densities and $\mathcal{L}_{\rm M}$
is the matter Lagrangian. In the above action because of the
topological origin of the Lovelock terms, the term proportional to
$c_i$ contributes to the equations of motion in dimensions with $n
\geq 2m$ , where $m$ is the order of Lovelock theory. Generally,
although the equations of motion of $k$th-order Lovelock gravity
are second-order differential equations, the $k$th-order Lovelock
term has no contribution to the field equations in $2k$ and lower
dimensions. for example the cubic term $\mathcal{L}_3$ does not
have any dynamical effect in five dimension. Recently, a
modification of Lovelock gravity called quasi-topological gravity
has been introduced, which has contribution to the field equations
in five dimensions from the $m$th-order ($m=3$) term in Riemann
tensor. Several aspects of $m$th-order quasi-topological terms
which have at most second-order derivatives of the metric in the
field equations for spherically symmetric space times in five and
higher dimensions except $2p$ dimensions have been investigated.

The action of $4$-th order quasi-topological gravity in ($n+1$)
dimensions can be written as follows
 \begin{equation}\label{Act2}
  I_{G}=\frac{1}{16\pi G_{n+1} }\int
  d^{n+1}x\sqrt{-g}[ {\mu}_{1}\mathcal{L}_{1}+{\mu} _{2}\mathcal{L}_{2}+{\mu}
_{3}\mathcal{X}_{3}+{\mu}
_{4}\mathcal{X}_{4}+\mathcal{L}_{\rm matter}],
  \end{equation}
which not only works in five dimensions but also yield
second-order equations of motion for spherically symmetric
spacetimes.

In action (\ref{Act2}),  $\mathcal{L}_{1}={R}$ is just the Einstein-Hilbert Lagrangian, $%
\mathcal{L}_{2}=R_{abcd}{R}^{abcd}-4{R}_{ab}{R}^{ab}+{R}^{2}$ is
the second order Lovelock (Gauss-Bonnet) Lagrangian and
\(\mathcal{L}_{\rm matter}\) is the Lagrangian of the matter
field. $\mathcal{X}_{3}$\ is the curvature-cubed Lagrangian given by
\cite{Myer1}
 \begin{eqnarray}
\mathcal{X}_{3} &=&R_{ab}^{cd}R_{cd}^{\,\,e\,\,\,f}R_{e\,\,f}^{\,\,a\,\,\,b}+%
\frac{1}{(2n-1)(n-3)}\left(
\frac{3(3n-5)}{8}R_{abcd}R^{abcd}R\right.  \notag
\\
&&-3(n-1)R_{abcd}R^{abc}{}_{e}R^{de}+3(n+1)R_{abcd}R^{ac}R^{bd}  \notag \\
&&\left. +\,6(n-1)R_{a}{}^{b}R_{b}{}^{c}R_{c}{}^{a}-\frac{3(3n-1)}{2}%
R_{a}^{\,\,b}R_{b}^{\,\,a}R+\frac{3(n+1)}{8}R^{3}\right) ,
\label{X3}
\end{eqnarray}%
 and $\mathcal{X}_4$ is the fourth order term of quasi-topological gravity \cite{MHD2}
 \begin{eqnarray}
\mathcal{X}_{4}\hspace{-0.2cm} &=&\hspace{-0.2cm}c_{1}R_{abcd}R^{cdef}R_{%
\phantom{hg}{ef}%
}^{hg}R_{hg}{}^{ab}+c_{2}R_{abcd}R^{abcd}R_{ef}R^{ef}+c_{3}RR_{ab}R^{ac}R_{c}{}^{b}+c_{4}(R_{abcd}R^{abcd})^{2}
\notag \\
&&\hspace{-0.1cm}%
+c_{5}R_{ab}R^{ac}R_{cd}R^{db}+c_{6}RR_{abcd}R^{ac}R^{db}+c_{7}R_{abcd}R^{ac}R^{be}R_{%
\phantom{d}{e}}^{d}+c_{8}R_{abcd}R^{acef}R_{\phantom{b}{e}}^{b}R_{%
\phantom{d}{f}}^{d}  \notag \\
&&\hspace{-0.1cm}%
+c_{9}R_{abcd}R^{ac}R_{ef}R^{bedf}+c_{10}R^{4}+c_{11}R^{2}R_{abcd}R^{abcd}+c_{12}R^{2}R_{ab}R^{ab}
\notag \\
&&\hspace{-0.1cm}%
+c_{13}R_{abcd}R^{abef}R_{ef}{}_{g}^{c}R^{dg}+c_{14}R_{abcd}R^{aecf}R_{gehf}R^{gbhd},
\label{X4}
\end{eqnarray}%
the coefficients $c_i$ in the above term are given by
\begin{eqnarray*}
c_{1} &=&-\left( n-1\right) \left( {n}^{7}-3\,{n}^{6}-29\,{n}^{5}+170\,{n}%
^{4}-349\,{n}^{3}+348\,{n}^{2}-180\,n+36\right) , \\
c_{2} &=&-4\,\left( n-3\right) \left( 2\,{n}^{6}-20\,{n}^{5}+65\,{n}^{4}-81\,%
{n}^{3}+13\,{n}^{2}+45\,n-18\right) , \\
c_{3} &=&-64\,\left( n-1\right) \left( 3\,{n}^{2}-8\,n+3\right) \left( {n}%
^{2}-3\,n+3\right) , \\
c_{4} &=&-{(n}^{8}-6\,{n}^{7}+12\,{n}^{6}-22\,{n}^{5}+114\,{n}^{4}-345\,{n}%
^{3}+468\,{n}^{2}-270\,n+54), \\
c_{5} &=&16\,\left( n-1\right) \left( 10\,{n}^{4}-51\,{n}^{3}+93\,{n}%
^{2}-72\,n+18\right) , \\
c_{6} &=&--32\,\left( n-1\right) ^{2}\left( n-3\right) ^{2}\left( 3\,{n}%
^{2}-8\,n+3\right) , \\
c_{7} &=&64\,\left( n-2\right) \left( n-1\right) ^{2}\left( 4\,{n}^{3}-18\,{n%
}^{2}+27\,n-9\right) , \\
c_{8} &=&-96\,\left( n-1\right) \left( n-2\right) \left( 2\,{n}^{4}-7\,{n}%
^{3}+4\,{n}^{2}+6\,n-3\right) , \\
c_{9} &=&16\left( n-1\right) ^{3}\left( 2\,{n}^{4}-26\,{n}^{3}+93\,{n}%
^{2}-117\,n+36\right) , \\
c_{10} &=&{n}^{5}-31\,{n}^{4}+168\,{n}^{3}-360\,{n}^{2}+330\,n-90, \\
c_{11} &=&2\,(6\,{n}^{6}-67\,{n}^{5}+311\,{n}^{4}-742\,{n}^{3}+936\,{n}%
^{2}-576\,n+126), \\
c_{12} &=&8\,{(}7\,{n}^{5}-47\,{n}^{4}+121\,{n}^{3}-141\,{n}^{2}+63\,n-9), \\
c_{13} &=&16\,n\left( n-1\right) \left( n-2\right) \left(
n-3\right) \left(
3\,{n}^{2}-8\,n+3\right) , \\
c_{14} &=&8\,\left( n-1\right) \left( {n}^{7}-4\,{n}^{6}-15\,{n}^{5}+122\,{n}%
^{4}-287\,{n}^{3}+297\,{n}^{2}-126\,n+18\right).
\end{eqnarray*}%

In the context of universal thermodynamics, our universe should be a
non-stationary gravitational system while from the cosmological
point of view, it should be homogeneous and isotropic. Therefore,
the natural choice is the FRW Universe, a dynamical spherically
symmetric spacetime, having only inner trapping horizon (the
apparent horizon), which is described by the line element
\begin{equation} \label{met1}
ds^2=h_{ab}dx^a dx^b+\tilde{r}^2 d \Omega ^2,
\end{equation}
where $x^0=t$, $x^1=r$, $\tilde{r}=a(t)r$, $a(t)$ is the scale
factor of the Universe with the curvature parameter $k$ with values
$-1, 0, 1$ corresponds to the open, flat and closed universes
respectively, $h_{ab}=diag(-1,a(t)^2/(1-kr^2))$ and $d \Omega^2$ is
the metric of $(n-1)$-dimensional unit sphere.

Varying the action (\ref{Act2}) with respect to metric leads to \cite{paper1}
\begin{equation}\label{Fri2}
\sum^m_{i=1}\hat{\mu}_i l^{
2i-2}\left(H^2+\frac{k}{a^2}\right)^{i}=\frac{16\pi
G_{n+1}}{n(n-1)}\rho.
\end{equation}
This is the Friedmann equation of arbitrary order
quasi-topological cosmology where $H$ is the Hubble parameter.
From now on, we sets $G_{n+1}=1$ for simplicity. Moreover,
$\hat{\mu}_i$'s are dimensionless parameters as follows
 \begin{eqnarray*}
   \hat{\mu}_{1}&=&1  \\
   \hat{\mu}_{2}&=&\frac{(n-2)(n-3)}{l^2}\mu_{2}  \\
   \hat{\mu}_{3}&=&\frac{(n-2)(n-5)(3 n^2-9n+4)}{8 (2n-1) l^4}\mu_3  \\
  \hat{\mu}_{4}&=&\frac{n(n-1)(n-2)^2(n-3)(n-7)(n^5-15 n^4+72 n^3-156
n^2+150 n-42)}{l^6}\mu_{4}.
\end{eqnarray*}
The dynamical apparent horizon, is determined by the relation
$h^{ab} \partial _a \tilde{r}\partial_b \tilde{r}=0$, where
$\tilde{r}=a(t) r$ . It is a matter of calculation to show that the
radius of the apparent horizon for the FRW universe is \cite{Hay2}
 \begin{equation}
\label{radius}
 \tilde{r}_A=\frac{1}{\sqrt{H^2+k/a^2}}.
\end{equation}
In cosmological context, various definitions of temperature are used
to get the corresponding Friedmann equations on the apparent
horizon. First, we use the original definition of temperature
(Hayward-Kodama temperature) together with the Clausius relation
($TdS_A = dQ^m$) as well as the unified form of the first law of
thermodynamics to extract an expression for the entropy of apparent
horizon in the quasi-topological cosmology. The Hayward-Kodama
temperature associated with the apparent horizon is defined as $T_h
= \kappa/2\pi$ where $\kappa$ is the surface gravity which can be
evaluated by using $\kappa
=\frac{1}{2\sqrt{-h}}\partial_{a}\left(\sqrt{-h}h^{ab}\partial_{b}\tilde
{r}\right)$ \cite{Bak,Cai3,Cao,GSL1,hel,hel1}. Therefore, the
surface gravity at the apparent horizon of the FRW universe has the
following form
 \begin{equation}\label{surgrav}
 \kappa=-\frac{1}{\tilde
r_A}\left(1-\frac{\dot {\tilde r}_A}{2H\tilde r_A}\right),
\end{equation}
which leads to
 \begin{equation}\label{temper}
 T_h=\frac{\kappa}{2\pi}=-\frac{1}{2 \pi \tilde
r_A}\left(1-\frac{\dot {\tilde r}_A}{2H\tilde r_A}\right),
\end{equation}
for the Hayward-Kodama temperature of apparent horizon. Then, using
the Cai-Kim temperature and the Clausius relation we get the entropy
of apparent horizon. In this approach, the horizon temperature is
\cite{Caikimt,Caikim}
\begin{equation}\label{caiii}
 T=\frac{1}{2 \pi\tilde{r}_A},
\end{equation}
and for an infinitesimal time interval, the horizon radius will have
a small change and we can use the $d\tilde{r}_A=0$ approximation
\cite{Caikimt,Caikim}.


\section{Unified first law of thermodynamics and horizon entropy }
\ \ The unified first law of thermodynamics can be expressed as \cite{Hay2, Hay3, Hay4}
\begin{equation}\label{UFL}
dE=A\Psi +WdV
\end{equation}
where $E$ is the total baryonic energy content of the
universe inside an $n$-sphere of volume $V$, while $A$ is the area
of the horizon. The energy flux $\Psi$ is termed as the energy
supply vector and $W$ is the work function which are
respectively defined as
\begin{equation}\label{wterm}
A\Psi=A(T_a^b \partial_b \tilde{r}_A+W\partial_a \tilde{r}_A),
\end{equation}
\begin{equation}\label{psiterm}
W=-\frac{1}{2}T^{ab}h_{ab}.
\end{equation}
The term $WdV$ in the first law comes from the fact that we have a
volume change for the total system enveloped by the apparent
horizon. For a pure de Sitter space, $\rho=−p$, and the work
term reduces to the standard $−pdV$, thus we obtain exactly the
standard first law of thermodynamics, $dE=TdS-pdV$. It is a
matter of calculation to show that Eq.~(\ref{wterm}) can be
written as
\begin{eqnarray}\label{UFLF1}
A\Psi=-AH\tilde{r}_A(\frac{\rho+p}{2})dt+Aa(\frac{\rho+p}{2})dr_A,
\end{eqnarray}
and thus
\begin{eqnarray}\label{UFLF}
A\Psi =-\frac{3V(\rho+p)H}{2}dt+\frac{A(\rho+p)}{2}(d\tilde{r}_A-\tilde{r}_A H dt),
\end{eqnarray}
on the apparent horizon of FRW universe. In obtaining the last
equation, we used $A\tilde{r}_A=3V$
relation.


\section{THERMODYNAMICS OF APPARENT HORIZON
IN QUASI-TOPOLOGICAL GRAVITY THEORY} Our Universe is undergoing an
accelerating expansion which represents a new imbalance in the
governing Friedmann equations. physicists have addressed such
imbalances by either introducing new sources or by changing the
governing equations. The standard cosmology model addresses this
imbalance by introducing a new source (dark energy) in the Friedmann
equations. On the contrary, a group of physicists have explored the
second route, i.e., a modified gravity approach, that at large
scales, Einstein theory of general relativity breaks down and a more
general action describes the gravitational field. In this study, we
follow the second approach.

\ \ The Friedmann equation of FRW Universe in quasi-topological
gravity represents by (\ref{Fri2}). In modified gravity theories,
the modified Friedmann equations can be written as
\begin{equation}\label{Fri3}
H^2+\frac{k}{a^2}=\frac{16\pi}{n(n-1)}\rho_t,
\end{equation}
and
\begin{equation}\label{Fri4}
\dot{H}-\frac{k}{a^2}=-\frac{8\pi
}{(n-1)}(\rho_t+p_t).
\end{equation}
In the above equations $\rho_t$ is the total energy density which
includes of non-interacting two fluid systems, one is the usual
fluid of energy density $\rho$ and thermodynamic pressure $p$,
while the second one is termed as effective energy density
$\rho_e$ due to curvature contributions and its corresponding
pressure $p_e$. So, we have
\begin{equation}
\rho_t = \rho + \rho_e  \ \ \ \ \ \ \ , \ \ \ \ \ \   \rho_t +p_t = (\rho+p) + (\rho_e+p_e)
\end{equation}
where
\begin{equation}\label{rhoef}
\rho_e =-\frac{n(n-1)}{16 \pi }\sum^m_{i=2}\frac{\hat{\mu}_i l^{
2i-2}}{\tilde{r}_A^{2i}}
\end{equation}
and
\begin{equation}\label{rhoef2}
\rho_e+p_e=-\frac{\varepsilon(n-1)}{4\pi }\sum^m_{i=2}\frac{i\hat{\mu}_i l^{2i-2}}{\tilde{r}_A^{2i}}
\end{equation}
where we have defined
$\varepsilon=\frac{\dot{\tilde{r}}_A}{2H\tilde{r}_A}$, in which
$\dot{\tilde{r}}_A=\frac{d\tilde{r}_A}{dt}$, and in terms of the
horizon radius $\tilde{r}_A$, we have \cite{cai9}
\begin{equation}\label{mod}
\dot{H}-\frac{k}{a^2}=-\frac{2\varepsilon}{\tilde{r}_A^2}.
\end{equation}
Finally, we should note that by combining Eqs.~(\ref{rhoef})
and~(\ref{rhoef2}) with each other, we get $\dot{\rho_e}+nH(\rho_e
+p_e) =0$. It means that Eqs.~(\ref{rhoef}) and~(\ref{rhoef2})
are only valid if the $\dot{\rho_e}+nH(\rho_e +p_e) =0$ and
$\dot{\rho}+nH(\rho+p) =0$ conditions be simultaneously available. These results
demonstrate that there is no energy exchange between the
geometrical and material fluids.
\subsection{Non-interacting case}
The terms due to the curvature, formally play the role of a
further source term in the field equations which effect is the
same as that of an effective fluid of purely geometrical origin.
 Indeed, since such terms modify the Einstein theory, they
lead to change the Bekenstein relation for the entropy of black
hole horizon \cite{Myer1,MHD2,MHD3,Brenn}, which is in agreement
with the result obtained by applying the first law of
thermodynamics on the black hole horizon \cite{DSH}. In order to
study the thermodynamics of apparent horizon of FRW universe in
quasi-topological gravity, authors in \cite{paper1} assumed that
the black hole entropy expression, derived in
\cite{Myer1,MHD2,MHD3,Brenn,DSH}, is also valid for the apparent
horizon of FRW universe. In addition, by applying the UFL of
thermodynamics on the apparent horizon they could get the
Friedmann equation in the quasi-topological gravity \cite{paper1}.
In fact, their recipe is a way of proposing an expression for the
apparent horizon entropy instead of a way for deriving the
corresponding entropy. Therefore, the inverse of their recipe may
indeed be considered theoretically as a more acceptable way of
getting the apparent horizon entropy. Finally, based on approach
\cite{paper1}, it seems that their proposal for the apparent
horizon entropy is independent of any interaction between the
cosmos sectors. It is also worthy to note that the mutual interaction between the geometrical ($T^e_{\mu \nu}$) and
non-geometrical ($T^m_{\mu \nu}$) sectors of cosmos in various
modified theories of gravity, affect the apparent horizon
entropy \cite{mitra,mitra1,mitra2,mitra3}. In fact, depending on
the scales, it is such a curvature fluid which may play the role
of dark energy. In Einstein general relativity framework, a dark
energy candidate and its interaction with other parts of cosmos
affects the Bekenstein entropy \cite{cana1,cana,em,mmg,mms}.
Therefore, the results obtained in
\cite{mitra,mitra1,mitra2,mitra3} are in line with those of
\cite{cana1,cana,em,mmg,mms}. In order to get the effects
of mutual interaction between the geometrical and non-geometrical
parts of cosmos, we need to establish a recipe to find out
the horizon entropy by starting from the Friedmann equations and
applying the UFL of thermodynamics on the apparent horizon of FRW
universe. The
energy conservation law leads to the continuity equation in the
form
\begin{equation}\label{contt}
\dot{\rho_t}+nH(\rho_t +p_t)=0.
\end{equation}
For a non interacting system it breaks down to
\begin{eqnarray}\label{contt1}
\dot{\rho}+nH(\rho +p) &=& 0 \\ \nonumber
\dot{\rho_e}+nH(\rho_e +p_e) &=& 0,
\end{eqnarray}
which implies that Eqs.~(\ref{rhoef}) and~(\ref{rhoef2}) are
available. Differentiating Eq. (\ref{Fri3}), we reach at
\begin{equation}\label{Fri4}
-\frac{2}{\tilde{r}_A^3}d\tilde{r}_A- \frac{16\pi}{n(n-1)}d\rho_e=\frac{16\pi}{n(n-1)}d\rho.
\end{equation}
Bearing Eq. (\ref{contt1}) in mind, we have
\begin{equation}\label{Fri5}
\frac{d\tilde{r}_A}{\tilde{r}_A^3}+ \frac{8\pi}{n(n-1)}d{\rho}_e=\frac{8\pi}{n-1}H \left(\rho+ p \right)dt.
\end{equation}
 Multiplying both sides of the above equation in ($-T$), one obtains
\begin{equation}\label{Fri6}
(-T)(\frac{d\tilde{r}_A}{\tilde{r}_A^3}+ \frac{8\pi}{n(n-1)}d{\rho}_e)=\frac{8\pi}{n-1}H \left(\rho+ p \right)dt\left[\frac{1}{2 \pi \tilde
r_A}\left(1-\frac{\dot {\tilde r}_A}{2H\tilde r_A}\right) \right].
\end{equation}
Assuming the total energy content of the universe inside an
$n$-sphere of radius $\tilde{r}_A$ is $E=\rho V$, where $
V=\Omega_n \tilde{r}_A^n$ is the volume enveloped by an
$n$-dimensional sphere. Taking differential form of the total
energy, after using the continuity equation~(\ref{contt1}), we
obtain
\begin{equation}\label{dE1}
dE=n\Omega_n \rho \tilde r_A^{n-1}d\tilde r_A-nH \Omega_n \tilde r_A^n(\rho +p)dt,
\end{equation}
which leads to
\begin{equation}\label{rhopdt}
(\rho +p)dt=-\frac{dE}{nH \Omega_n \tilde r_A^n}+ \frac{\rho d\tilde r_A}{H\tilde r_A}.
\end{equation}

Substituting Eq.~(\ref{rhopdt}) into~(\ref{Fri6}), simple
calculations leads to
 \begin{equation}\label{Frid7}
T(\frac{d\tilde{r}_A}{\tilde{r}_A^3}+ \frac{8\pi}{n(n-1)}d{\rho}_e)=\frac{4dE}{n(n-1)\Omega_n \tilde r_A^{n+1}}-\frac{2(\rho-p)d\tilde r_A}{(n-1)\tilde r_A^2},
\end{equation}
and consequently,
 \begin{equation}\label{Frid8}
T(\frac{d\tilde{r}_A}{\tilde{r}_A^3}+ \frac{8\pi}{n(n-1)}d{\rho}_e)=\frac{4}{n(n-1)\Omega_n \tilde r_A^{n+1}}\left[dE-WdV\right],
\end{equation}
where we have used $dV=n \Omega_n \tilde r_A^{n-1}d\tilde r_A$. In
this equation the work density $W=\frac{\rho-p}{2}$ is regarded as
the work done when the apparent horizon radius changes from
$\tilde{r}_A$ to $\tilde{r}_A+d\tilde{r}_A$. The Clausius relation
is \cite{em}
\begin{equation}\label{ent}
T dS_A=\delta Q^m=A \Psi,
\end{equation}
where $\delta Q^m$ is the energy flux crossing horizon during the
universe expansion. Eqs.~(\ref{UFL}), (\ref{Frid8}) and (\ref{ent}),
we have
\begin{equation}\label{dsquasi}
dS_A=\frac{n(n-1)\Omega_n \tilde r_A^{n-2}d\tilde r_A}{4}+2 \pi \Omega_n \tilde r_A^{n+1}d\rho_e.
\end{equation}
Using Eq.~(\ref{contt1}) leads to
\begin{equation}\label{dsquasi00}
dS_A=\frac{n(n-1)\Omega_n \tilde r_A^{n-2}d\tilde r_A}{4}-2nH \pi \Omega_n \tilde r_A^{n+1}(\rho_e+p_e)dt.
\end{equation}
Now, by either inserting Eq.~(\ref{rhoef}) into Eq.~(\ref{dsquasi})
or Eq.~(\ref{rhoef2}) into Eq.~(\ref{dsquasi00}), and integrating
the result, we get
\begin{equation}\label{dsquasi2}                                                                                                                                                                                                                                                                                                                                                                                                                                                                                                                                                                               S_A=\frac{A}{4}\sum^{m}_{i=1}  i
\frac{(n-1)}{(n-2i+1)}\frac{\hat{\mu}_{i}l^{2i-2}}{\tilde{r}^{2i-2}_{A}}+S_0,
\end{equation}
for the horizon entropy. This result is in full agreement with
previous proposal in which authors assumed that the event horizon
entropy is extendable to the apparent horizon of FRW universe
\cite{paper1}. $S_0$ is an integration constant and we can consider
it zero without lose of generality. It is also useful to mention
here that we considered the Clausius relation in the form
$TdS_A=\delta Q^m$ to obtain this result, while authors in
\cite{paper1} took into account the $TdS_A=-\delta Q^m$ form of the
Clausius relation. This discrepancy is due to our different way of
defining the horizon temperature. We have used the Hayward-Kodama
definition of temperature~(\ref{temper}), while authors in
\cite{paper1} used its absolute value to avoid attributing negative
temperatures on the apparent horizon and thus the Hawking radiation.

\subsection*{The Cai-Kim approach}
Here, we focus on the Cai-Kim approach to get the effects of
geometrical fluid on the horizon entropy. When one applies the first
law on the apparent horizon to calculate the surface gravity and
thereby the temperature and considers an infinitesimal amount of
energy crossing the apparent horizon, the apparent horizon radius
$\tilde{r}_A$ should be regarded to have a fixed value ($dV=0$)
\cite{Caikim}. Bearing Eq.~(\ref{UFLF}) together with the
$d\tilde{r}_A=0$ approximation in mind, by using Eq.~(\ref{contt1})
and the Clausius relation in the form $TdS_A=-\delta Q^m=-A\Psi$,
where $\delta Q^m$ is the energy flux crossing the horizon during
the infinitesimal time interval $dt$, we get
\begin{equation}\label{Caikim0}
TdS_A=-Vd\rho.
\end{equation}
Substituting $d\rho$ from Eq.~(\ref{Fri4}) into the above equation, we are led to
\begin{equation}\label{Caikim2}
dS_A=-\frac{V}{T}d\rho =2\pi \Omega_n \tilde{r}_A^{n+1}\left[\frac{n(n-1)d\tilde{r}_A}{8\pi \tilde{r}_A^3}+d\rho_e\right].
\end{equation}
where we have used $V=\Omega_n\tilde{r}_A^n$ and the Cai-Kim temperature ($T=\frac{1}{2\pi\tilde{r}_A}$) \cite{Caikimt,Caikim}. Finally, we obtain
\begin{equation}\label{dsquasi3}
dS_A=\frac{n(n-1)\Omega_n \tilde r_A^{n-2}d\tilde r_A}{4}+2 \pi \Omega_n \tilde r_A^{n+1}d\rho_e.
\end{equation}
Consequently the modified entropy on the event horizon has the explicit form
\begin{equation}\label{dsquasi4}                                                                                                                                                                                                                                                                                                                                                                                                                                                                                                                                                                                                                              S_A=\frac{A}{4}\sum^{m}_{i=1}  i
 \frac{(n-1)}{(n-2i+1)}\frac{\hat{\mu}_{i}l^{2i-2}}{\tilde{r}^{2i-2}_{A}}+S_0.
\end{equation}
Which is compatible with the results obtained by considering the
Hayward-Kodama temperature (\ref{dsquasi2}). As before, since
entropy is not an absolute quantity, without lose of generality we
can set $S_0$ to zero. Setting $n=3$ and $i=1$ in the above result,
leads to the Bekenstein-Hawking entropy formula.
\subsection{Interacting universe}

In order to take the interaction into account, for the
quasi-topological gravity, consider the energy-momentum
conservation law in the forms
\begin{equation}\label{intera1}
\dot{\rho}+nH(\rho +p)=Q
\end{equation}
 and
 \begin{equation}\label{intera2}
\dot{\rho_e}+nH(\rho_e +p_e)=-Q,
\end{equation}
where $Q$ is the mutual interaction between different parts of the
cosmos. The latter means that Eqs.~(\ref{rhoef}) and~(\ref{rhoef2})
are not simultaneously valid for an interacting universe. Using
Eqs.~(\ref{Fri4}) and~(\ref{intera1}), we get
 \begin{equation}\label{diff1}
\frac{d\tilde{r}_A}{\tilde{r}_A^3}+\frac{8\pi }{n(n-1)}(d\rho_e+Qdt)=\frac{8\pi H}{n-1}(\rho+p)dt.
\end{equation}
Multiplying both sides of the above equation by $(-T)$, one obtains
 \begin{equation}\label{diff2}
T\left(\frac{d\tilde{r}_A}{\tilde{r}_A^3}+\frac{8\pi }{n(n-1)}(d\rho_e+Qdt)\right)=-\frac{2 H(\rho +p)}{(n-1)\tilde{r}_A}dt+\frac{2(\rho+p)}{(n-1)\tilde{r}_A^2} adr_A,
\end{equation}
where we have used $\dot{\tilde{r}}_Adt=adr_A+r_A H dt$.
Multiplying the result by the factor $\frac{n(n-1)\Omega_n
\tilde{r}_A^{n+1}}{4}$, we arrive at
 \begin{equation}\label{diff3}
T\left(\frac{d\tilde{r}_A}{\tilde{r}_A^3}+\frac{8\pi
}{n(n-1)}(d\rho_e+Qdt)\right)(\frac{n(n-1)\Omega_n
\tilde{r}_A^{n+1}}{4})=n\Omega_n\tilde{r}_A^{n}[-\frac{\rho
+p}{2}Hdt+\frac{\rho+p}{2\tilde{r}_A} adr_A].
\end{equation}
Simple calculations leads to
\begin{equation}\label{diff4}
T\left(\frac{n(n-1) \Omega_n \tilde{r}_A^{n-2}d\tilde{r}_A}{4}+2\pi \Omega_n \tilde{r}_A^{n+1}(d\rho_e+Qdt)\right)=-AH\frac{\rho +p}{2}\tilde{r}_A dt+Aa\frac{\rho+p}{2}dr_A.
\end{equation}
The right hand side of Eq. (\ref{diff4}) is nothing but the energy
flux ($\delta{Q}^m$) crossing the apparent horizon~(\ref{UFLF1}).
Bearing the Clausius relation~(\ref{ent}) in mind, we get
\begin{equation}\label{diff5}
dS_A=\frac{n(n-1) \Omega_n \tilde{r}_A^{n-2}d\tilde{r}_A}{4}+2\pi \Omega_n \tilde{r}_A^{n+1}(d\rho_e + Qdt).
\end{equation}
Now, using Eq.~(\ref{intera2}) to obtain
\begin{equation}\label{diff6}
dS_A=\frac{n(n-1) \Omega_n \tilde{r}_A^{n-2}d\tilde{r}_A}{4}-2\pi nH\Omega_n \tilde{r}_A^{n+1}(\rho_e +p_e)dt,
\end{equation}
which leads to
\begin{equation}\label{diff7}
S_A=\frac{A}{4}-2\pi n\Omega_n\int H\tilde{r}_A^{n+1}(\rho_e +p_e)dt+S_0,
\end{equation}
which is in agreement with the results obtained by authors
\cite{mitra} for the Lovelock theory but in a different way. In
fact, since in the interacting universe $\rho_e$ and $\rho_e+p_e$
cannot simultaneously meet Eqs.~(\ref{rhoef}) and~(\ref{rhoef2}),
respectively, we can not use them to integrate from the RHS of
Eqs.~(\ref{diff5}) and~(\ref{diff6}) to get an expression for the
horizon entropy. Now, consider a special situation in which $\rho_e$
satisfies Eq.~(\ref{rhoef}), from Eq.~(\ref{diff5}) we get
\begin{equation}\label{diff5c}
dS_A=\frac{n(n-1) \Omega_n \tilde{r}_A^{n-2}d\tilde{r}_A}{4}+2\pi \Omega_n \tilde{r}_A^{n+1}d\rho_e+ 2\pi \Omega_n \tilde{r}_A^{n+1}Qdt.
\end{equation}
In the above equation, first two terms of the right hand side is
exactly the same as Eq.~(\ref{dsquasi3}). In other words,
\begin{equation}\label{entint}
(S_A)_{interaction}=(S_A)_{non-interaction}+2\pi \Omega_n \int \tilde{r}_A^{n+1}Qdt.
\end{equation}
It is crystal clear that the second term of RHS of this equation
counts the effects of mutual interaction between the geometrical
and non-geometrical fluids on the horizon entropy. Loosely
speaking, the above expression for horizon entropy is no longer
the usual Bekenstein entropy formula, rather there are two
correction terms which are due to higher order curvature terms,
arising from the quasi-topological nature of gravity theory, along
with the mutual interaction between the geometrical and
non-geometrical parts. Note that in this situation, due to
equation  Eq.~(\ref{intera2}), the $\rho_e+p_e$ term does not meet
Eq.~(\ref{rhoef2}). As another example, consider a situation in
which $\rho_e+p_e$ satisfies Eq.~(\ref{rhoef2}), which also means
that, due to equation  Eq.~(\ref{intera2}), $\rho_e$ does not obey
Eq.~(\ref{rhoef}). For this case, by combining
Eqs.~(\ref{intera2}) and~(\ref{diff5}), and integrating the result, we obtain
\begin{equation}
(S_A)_{interaction}=(S_A)_{non-interaction},
\end{equation}
which means that the mutual interaction $Q$ does not affected the
horizon entropy in this special case.

\subsection*{The Cai-Kim approach}
The rest of this section devotes to achieve Eq. (\ref{diff6})
by using the Cai-Kim approach \cite{Caikimt,Caikim}. For an
interacting universe the UFL is in the form \cite{mms}
\begin{equation}\label{cai22}
dS_A=-\frac{V}{T}(d\rho-Qdt).
\end{equation}
Using Eqs. (\ref{Fri4}) and (\ref{intera2}), we have
\begin{equation}\label{cai23}
d\rho-Qdt=-\frac{n(n-1)d\tilde{r}_A}{8\pi \tilde{r}_A^3} +nH(\rho_e+p_e)dt.
\end{equation}
Substituting it into Eq. (\ref{cai22}), leads to
\begin{equation}\label{cai24}
dS_A=\frac{n(n-1) \Omega_n \tilde{r}_A^{n-2}d\tilde{r}_A}{4}-2\pi nH\Omega_n \tilde{r}_A^{n+1}(\rho_e +p_e)dt,
\end{equation}
which is in full agreement with Eq. (\ref{diff6}).
\section{second law of thermodynamics}\label{2thlow}
The time evolution of the entropy in a non-interacting universe
governed by quasi-topological gravity, is already considered by
authors in \cite{paper1}. Now, we are interested in examining the
validity of the second law of thermodynamics in an interacting
universe. This low states that the horizon entropy should meet the
$\frac{dS_A}{dt}\geq0$ condition \cite{haw}. For this propose, from
Eq. (\ref{diff6}) we have
\begin{equation}\label{dia}
\frac{dS_A}{dt}=\frac{n(n-1) \Omega_n \tilde{r}_A^{n-2}\dot{\tilde{r}}_A}{4}-2\pi nH\Omega_n \tilde{r}_A^{n+1}(\rho_e +p_e).
\end{equation}
However, using Eqs.~(\ref{intera2}) and~(\ref{diff1}), one gets
\begin{equation}\label{diff1a}
\frac{\dot{\tilde{r}}_A}{\tilde{r}_A^3}-\frac{8\pi H}{n-1}(\rho_e+p_e)=\frac{8\pi H}{n-1}(\rho+p),
\end{equation}
which leads to the Raychaudhuri equation
\begin{equation}\label{Raych}
\dot{\tilde{r}}_A=\frac{8\pi H}{n-1}(\rho+p+\rho_e+p_e)\tilde{r}_A^3.
\end{equation}
Substituting it into Eq. (\ref{dia}), we arrive at
\begin{equation}\label{2thlow2}
\frac{dS_A}{dt}=2\pi nH\Omega_n \tilde{r}_A^{n+1}(\rho+p).
\end{equation}
The above result indicate that the second law of thermodynamics
($\frac{dS_A}{dt}\geq0$) is valid for the apparent horizon under the
condition $(\rho+p)>0$ and if we define the state parameter
$\omega=\frac{p}{\rho}$, this law is valid whenever the state
parameter obeys $\omega\geq-1$.
\section{Summary and Conclusions\label{con}}
After giving a brief review of the quasi-topological gravity
theory, by considering a FRW universe, we pointed out to the
corresponding Friedmann equation in this modified gravity theory.
In section ($\textsc{II}$), we mentioned the apparent
horizon of FRW universe as the proper causal boundary, its surface
gravity and the Hayward-Kodama temperature of this hypersurface.
Motivated by recent works on the thermodynamics of horizon of FRW
universe in some modified gravity theories
\cite{mitra,mitra1,mitra2,mitra3}, we considered the terms other than
Einstein tensor as a fluid with energy density $\rho_e$  and pressure
$p_e$. In fact, since these terms may
play the role of dark energy and because a dark energy candidate
may also affect the horizon entropy \cite{cana,cana1,em,mmg,mms},
such justification at least is not forbidden. Then, we showed
that if we either use the Hayward-Kodama or the Cai-Kim
temperature, the same result for the horizon entropy is obtainable
(Eqs.~(\ref{dsquasi2}) and~(\ref{dsquasi4})) which is in agreement
with previous work \cite{DSH}, in which authors by using different
definition for the horizon temperature and generalizing the black
hole entropy to the cosmological horizon setup, could propose the
same expression as that of us for the apparent horizon entropy in
quasi-topological gravity. Moreover, it is useful to note
that Eq.~(\ref{dsquasi2}) is valid only if there is no
energy-momentum exchange between the geometrical ($T^e_{\mu \nu}$)
and non-geometrical ($T^m_{\mu \nu}$) fluids.

Finally, we generalized our investigation to an interacting FRW
universe with arbitrary curvature parameter ($k$) in
quasi-topological gravity. We got a relation for the horizon
entropy (Eq.~(\ref{diff7})) which showed the effects of an
energy-momentum exchange between the geometrical and
non-geometrical fluids on the apparent horizon entropy.
Thereinafter, we studied the special case in which $\rho_e$ meets
Eq.~(\ref{rhoef}) to perceive clearly the effects of such mutual
interaction on the horizon entropy. We got Eq.~(\ref{entint}) for
the horizon entropy, which showed two terms besides the Bekenstein
entropy due to higher order curvature terms, arising from the
quasi-topological nature of model, and mutual interaction between
the geometrical and non-geometrical fluids. Our investigation also
showed that whenever $\rho_e+p_e$ satisfies Eq.~(\ref{rhoef2}), the
mutual interaction $Q$ does not affect the horizon entropy and
therefore, the result of non-interacting case is obtainable. We
have also pointed out the validity of second law of thermodynamics
in an interacting case and found out that, independent of
curvature parameter ($k$), whenever $\rho+p\geq0$, the second law
is obtainable.

Our study shows that, in quasi-topological gravity theory, a
geometrical fluid with energy density $\rho_e$ and pressure $p_e$,
independent of its nature, affects the apparent horizon entropy as
\begin{equation}
S_A=\frac{A}{4}-2\pi n\Omega_n\int H\tilde{r}_A^{n+1}(\rho_e +p_e)dt,
\end{equation}
which is in full agreement with other studies
\cite{cana,cana1,em,mmg,mms,mitra,mitra1,mitra2,mitra3}. Since such
correction in quasi-topological gravity has geometrical nature, it
is indeed inevitable. Moreover, the effects of mutual interaction
between the dark energy candidate and other parts of cosmos is
stored in the second term of RHS of this equation. Indeed, based on
our approach, one may find the apparent horizon entropy of FRW
universe with arbitrary curvature parameter ($k$) in modified
theories of gravity by interpreting the terms other than Einstein
tensor in the corresponding Friedmann equation as a geometrical
fluid and applying the UFL of thermodynamics on the apparent
horizon. At the end, we should stress that in the absence of a
mutual interaction, the results of interacting universes in
quasi-topological theory converge to those of the non-interacting
case.
\acknowledgments{This work has been supported financially by Research Institute for Astronomy \& Astrophysics of Maragha (RIAAM), Iran.}

\end{document}